\documentclass[twocolumn,showpacs,preprintnumbers,superscriptaddress,amsmath]{revtex4}
\usepackage[dvips]{graphics}
\usepackage{epsfig}
\usepackage{url}
\usepackage{psfrag}
\usepackage[psamsfonts]{amssymb}
\usepackage{amsmath}
\usepackage{indentfirst}
\usepackage{amssymb}
\usepackage{wrapfig}
\usepackage{url}
\usepackage{dcolumn}
\usepackage{bm}

\newcommand{\be}{\begin{equation}}
\newcommand{\ee}{\end{equation}}
\newcommand{\ba}{\begin{eqnarray}}
\newcommand{\ea}{\end{eqnarray}}

\begin{document}

\preprint{APS preprint}

\title{Generic Multifractality in Exponentials of Long Memory Processes}


\author{A. Saichev}
\affiliation{Mathematical Department, Nizhny Novgorod
State University, Gagarin prosp. 23, Nizhny Novgorod,
603950, Russia} \affiliation{Institute of Geophysics and
Planetary Physics, University of California, Los Angeles,
CA 90095}

\author{D. Sornette}
\affiliation{Institute of Geophysics and Planetary
Physics and Department of Earth and Space Sciences,
University of California, Los Angeles, CA 90095}
\affiliation{Laboratoire de Physique de la Mati\`ere
Condens\'ee, CNRS UMR 6622 and Universit\'e de
Nice-Sophia Antipolis, 06108 Nice Cedex 2, France}
\email{sornette@moho.ess.ucla.edu}

\date{\today}

\begin{abstract}

We find that multifractal scaling is a robust property of a large class of
continuous stochastic processes, constructed as exponentials of
long-memory processes.  The long memory is characterized
by a power law kernel with tail exponent
$\varphi+1/2$, where $\varphi >0$. This generalizes previous studies performed
only with $\varphi=0$ (with a truncation at an integral scale), by showing
that multifractality holds over a remarkably large range of 
dimensionless scales for $\varphi>0$. The intermittency multifractal coefficient can
be tuned continuously as a function of the deviation $\varphi$ from $1/2$
and of another parameter $\sigma^2$
embodying information on the short-range amplitude of the memory kernel, 
the ultra-violet cut-off (``viscous'') scale and the variance of the white-noise innovations.
In these processes, both a viscous scale and an integral scale naturally appear, bracketing
the ``inertial'' scaling regime. We exhibit a surprisingly good collapse of the
multifractal spectra $\zeta(q)$ on a universal scaling function, which enables
us to derive high-order multifractal exponents from the small-order values
and also obtain a given multifractal spectrum $\zeta(q)$ 
by different combinations of $\varphi$ and $\sigma^2$.

\end{abstract}


\maketitle

\section{Introduction}

Generalizing the cascade models that
started with Richardson \cite{rich} and Kolmogorov \cite{kolmo}, 
multifractal cascades have been introduced in turbulence \cite{M1,M2}
to model the anomalous scaling exhibited 
by the moments of the velocity increments in hydrodynamic turbulence (see
for instance \cite{Frisch} and references therein). They have been since
applied to many other complex fields including fractal growth processes,
geophysical fields, high energy particle physics, astronomy, biology and
finance \cite{BB}. The constructions involved in multifractal 
cascades are based on hierarchical geometries coupled with multiplicative
noise and form discrete hierarchical cascades \cite{jmp}. They have been very
useful to highlight a general mechanism for intermittency and
multifractality which reflects the presence of intermittent bursts of
fluctuations with long-range correlations. Accordingly, the long-range
correlations are seen to result from the large-scale structures that
impact the smaller scales through a hierarchical cascade. But
discrete cascades have limitations and defects such as spurious effects
due to the discreteness (scaling holds
only for certain scale ratios \cite{dsi}), non-stationarity, and 
absence of causality in the time domain (see however \cite{aArn98}).

Here, we study the multifractal properties of a class of continuous
stochastic processes, constructed as exponentials of long-memory
processes with power law memory. Previous works briefly reviewed below
have been concerned with the case where the power law memory has a tail
exponent equal to $1/2$, which leads to a logarithmically decaying
correlation functions and the necessity for a regularization at large
time scales, i.e., the introduction of a so-called integral scale. With
logarithmic correlation functions, previous works have shown that
this class of processes exhibit the property of multifractality.
Here, we extend the problem to power law memory with tail exponent $\varphi+1/2$
which can vary arbitrarily above $1/2$. As a consequence of the 
faster-than-logarithmic decay of the correlation function of the process, 
the property of multifractality cannot hold exactly anymore. We show
however that multifractality holds over a remarkably large range of 
dimensionless scales and that the intermittency coefficients can
be tuned continuously as a function of the deviation $\varphi$ from $1/2$
of the exponent of the power law memory and of another parameter $\sigma^2$
embodying information on the short-range amplitude of the memory kernel, 
the ultra-violet cut-off scale and the variance of the white-noise innovations.
For this, we present a motivated robust algorithm to determine
the exponents $\zeta(q)$ of the multifractal spectrum that we apply
on our numerically determined structure functions or moments of the
stochastic process. We exhibit a surprisingly good collapse of the
multifractal spectra on a universal scaling function, which enables
us to derive high-order multifractal exponents from the small-order values.
The scaling ansatz is validated by direct numerical evaluations
of integral expressions of the moments of orders up to $q=5$.
Our results offer an interesting generalization of the class 
of multifractal random walks introduced recently and provide a
physically interpretable source of multifractal intermittency in 
terms of the parameters $\varphi$ and $\sigma^2$.
Our results have potential use in all the fields in which multifractal
properties have been discussed in the time domain. For instance, 
they provide a rational for the approximate multifractal signatures
observed in simple agent-based models of social networks \cite{Zhousocial},
without the need to justify an exact logarithmic scaling for the
correlation function of the logarithm of the observable.

The organization of the paper is the following.  Section 2 presents
a short review to position the stochastic process which we also define.
Section 3 defines the effective multifractal exponents and present
a simple efficient method for their estimation. Section 4 focuses
on the scaling properties of the second-order moment and its associated
exponent $\zeta(2)$. Section 5 presents our main results for the higher-order
moments and their exponents $\zeta(q)$. Section 6 offers some numerical
and mathematical insights on the origin of the effective multifractality.
Section 7 concludes.

\section{Stochastic continuous processes as exponentials
of processes with power law memory with arbitrary exponents}

Recently, inspired by the logic of the construction of discrete
hierarchical cascades, several works have divised genuine stochastic
continuous stationary processes which reproduce their main properties
\cite{BacriDelourMuzy,Schmitt1,Muzybacry,SMM,BM03,Schmitt2,Barral1}.
In particular, the so-called multifractal random walk (MRW) has been
introduced by Bacry, Delour and Muzy as the only continuous stochastic 
stationary causal process
with exact multifractal properties and Gaussian infinitesimal increments
\cite{BacriDelourMuzy}. Sornette et al. \cite{SMM} have shown that 
the increments $\delta_\tau X(t) $ at finite scale $\tau$ of
the MRW can be approximated by
\be
\delta_\tau X(t) =  \int_{t-\tau}^t \epsilon(t) \cdot e^{\omega_{\tau}(t)}~,
\label{remglww}
\ee
where $\delta_\tau X(t)= X(t+\tau)- X(t)$, 
$\epsilon(t)$ is a standardized Gaussian
white noise independent of $\omega_{\tau}(t)$ and
$\omega_{\tau}(t)$ can be expressed as 
an auto-regressive process
\be
\omega_{\tau}(t) = \omega^0_{\tau}+\int_{-\infty}^t dW(t')~
h_{\tau}(t-t')~,
\label{mbhmle}
\ee
where $W(t)$ denotes a Wiener process with unit diffusion coefficient and
the memory kernel $h_{\tau}(\cdot)$ is a causal function
specified by its Fourier transform 
\be
[\hat h_{\tau}(f)]^2  = 2 \lambda^2~
f^{-1}\left[\int_0^{Tf}\frac{\sin(t)}{t} dt+O\left(f \tau
\ln(f \tau)\right)\right]~.
\label{mgmkala}
\ee
The expression (\ref{mgmkala}) shows that
\be
h_{\tau}(t') \sim K_0 \sqrt{\frac{\lambda^2 T}{t'}} ~~~~~
\mbox{for}~~ \tau \ll t' \ll T~,
\label{mgmlww}
\ee
where the so-called integral scale $T$ delineates the boundary 
beyond which the correlation vanishes exactly.
This slow inverse square-root power law decay (\ref{mgmlww})
of the memory kernel in (\ref{mbhmle})
ensures the long-range logarithmic dependence of the
correlation function of $\omega_{\tau}(t)$ \cite{SMM}, which is one
important ingredient for the multifractality of $\delta_{\tau} X(t)$.
Schmitt \cite{Schmitt2} has studied in details the stochastic
process (\ref{remglww},\ref{mbhmle}) with a kernel $h_{\tau}(t')$
exactly given by the square-root power law (\ref{mgmlww})
for $t'$ between scale $\tau =1$ and $T$, with a smooth regularization
for $t < \tau$. Not surprisingly, this process exhibits
multifractality in the range of scales between $\tau$ and $T$.

We study the positive stochastic process $\delta_{\tau} X(t)$
generalizing (\ref{remglww},\ref{mbhmle}) with (\ref{mgmlww}), defined by
\be
\delta_\tau X(t)= \int_{t-\tau}^{t} \mu(t') dt'~ , ~~~{\rm with}~ \mu(t) = \kappa~e^{\omega(t)}~,
\label{5}
\ee
with $\omega(t)$ of the form (\ref{mbhmle}) 
\be
\omega(t) = \int_{-\infty}^t dW(t')~ h(t-t')~,
\label{mbhmle2}
\ee
and
\be
h(t)= {h_0 \over (1+x)^{\varphi+1/2}} H(t)~ , \qquad x= t/ \ell~ ,
\label{10}
\ee
where $\ell$ is some ``microscopic'' characteristic scale, regularizing the
singularity of the power law in the propagator $h(t)$ at $t=0$ and $H(t)$ is the unit
step Heaviside function ensuring the condition of causality inherent to
most applications. The main departure from the previously cited works is
to consider an exponent ${1 \over 2}+\varphi$ larger that $1/2$ for the 
power law decay of the memory kernel $h(t)$. As already
mentioned, one can prove rigorously \cite{BM03} that the process 
(\ref{remglww},\ref{mbhmle}) with (\ref{mgmkala}), whose power law
approximation for the memory kernel has the exact exponent $1/2$, 
i.e. $\varphi=0$, has a logarithmic decaying
covariance function of the auxiliary stationary Gaussian
stochastic process $\omega(t)$
\be
\langle \omega(t) \omega(t+\tau)\rangle\sim \ln\left({T \over
\tau}\right)~ , \qquad \tau<T~ , 
\label{7}
\ee
associated with the multifractal signature of the process $X(t)$ given by
\be
\langle [\delta_\tau X(t)]^q \rangle= a(q) \tau^{\zeta(q)}~ , ~~~{\rm for}~\tau<T~.
\label{2}
\ee
In expression (\ref{2}), the angle brackets denote a statistical averaging
and the multifractal ``spectrum'' $\zeta(q)$ has the parabolic form
\be
\zeta(q)= \left(1+ {\lambda^2 \over 2}\right) q- {\lambda^2 \over 2}
q^2~ , 
\label{8}
\ee
where $\lambda^2=-\zeta''(0)$ is the so-called \emph{intermittency
coefficient}. 
In contrast, the existence of multifractality defined by the
nonlinear spectrum $\zeta(q)$ has not been studied
previously for the process (\ref{5},\ref{10}) with non-zero 
values $\varphi>0$.
It is not obvious a priori that multifractality will be observed
because the deviation from $1/2$ for the exponent of the power
law decay of the memory kernel implies that the covariance function of the
stochastic process $\omega(t)$ is no more logarithmic, which
was the fundamental reason for the existence of multifractality in
the MRW. However, Ouillon and Sornette \cite{PRL05,JGR05} have recently
shown that the process (\ref{5},\ref{10}) with non-zero values $\varphi>0$ has
robust multi-scaling properties. They have derived this process from 
the physics of thermally activated rupture and long memory
stress relaxation for earthquakes, and have shown that this process
predicts that seismic decay rates after mainshocks follow approximately the 
Omori law $\sim 1/t^p$ with exponents $p(M)$ linearly increasing with the 
magnitude $M$ of the mainshock, in agreement with observations \cite{PRL05,JGR05}. 
Such multi-scaling suggest that the property of
multifractality in the sense of (\ref{2}) should also be present. 

A significant difference between the process (\ref{5},\ref{10}) with
non-zero values $\varphi>0$ and with $\varphi=0$ is that no integral scale $T$
is needed to regularize the theory. Furthermore, all moments 
$\langle [\delta_\tau X(t)]^q \rangle$ are finite for $\varphi>0$
whereas all the moments of order $q>q_*$ for $\varphi=0$, where $q_*$ satisfies to equation
$\zeta(q_*)=1$, are infinite. This divergence signals that the 
probability density function (PDF) of the increments $\delta_\tau X(t)$
is heavy-tailed with an exponent smaller than $q_*$ \cite{Muzytail}.
The absence of divergence of all moments of the PDF for $\varphi>0$ excludes a heavy tail
but not fat tails of the PDF of increments, such as stretched exponentials (which are known
to approach arbitrary well any power law, see chapter 2 of \cite{MalSor}).
Thus, if multifractality exists for $\varphi>0$, it may be observed
for any $q \geq 0$. 

In addition to $\varphi$, the other key parameter controlling the multifractal
properties of the process (\ref{5},\ref{10}) will be shown to be
\be
\sigma^2= \int_0^\infty h^2(t) dt = h_0^2 { \ell \over 2\varphi}~.
\label{11}
\ee
Note the divergence of $\sigma^2$ for $\varphi \to 0$, for which the integral
scale must be re-introduced to regularize the theory.

\section{Definition and determination of effective multifractal exponents}

\subsection{Definitions and notations}

A general theoretical characterization of the increments (\ref{5}) is 
offered by the moment functions defined by
\be
M(t_1,\dots,t_q)= \langle \prod_{r=1}^q \mu(t_r)\rangle= \langle
\mu^q\rangle \prod_{i=1 \atop k=i+1}^q G(x_k-x_i) 
\label{gjjaclma}
\ee
of the lognormal density $\mu(t) = \kappa~e^{\omega(t)}$ defined
in (\ref{5}). In (\ref{gjjaclma}), we use the following notations: $x_i=t_i/\ell$,
\be
\langle \mu^q\rangle= \kappa^q e^{\sigma^2 q^2/2}~ , ~~ G(y)=
e^{-\sigma^2 d(y)}~ , ~~ d(y)= 1- C(y)~ , 
\label{12}
\ee
where $C(t)$ is the normalized ($C(0)=1$) covariance function of the Gaussian process $\omega(t)$ 
defined in (\ref{mbhmle2}),
\be
C\left({\tau\over\ell}\right)= {1 \over \sigma^2} \int_0^\infty
h(t)h(t+\tau) dt~.
\label{nbjvnvqw}
\ee

We start our investigation of the multifractal properties of the
increments $\delta_\tau X(t)$ defined in (\ref{5}) by calculating
the moment $\langle [\delta_\tau X(t)]^q \rangle$ for $q=2$ and checking
if the power law scaling of the form (\ref{2}) holds with an
exponent $\zeta(2)$ smaller than $2$. If this is the case, 
and from the fact that $\zeta(0)=0$ and $\zeta(1)=1$, we can conclude
that $\zeta(q)$ is a nonlinear function of $q$, the hallmark of 
multifractality. The fact that $\zeta(1)=1$ results from the
positivity of the stationary density $\mu(t)$.

In order to get the exponent $\zeta(2)$, let us study the normalized second moment 
\be
S_2(y)= {1 \over \ell^2 \langle\mu^2\rangle}\langle
\left[\delta_\tau X(t)\right]^2\rangle= \int_0^y dx_1 \int_0^y dx_2~
G(x_2-x_1)~ .
\label{jgjjbnwb}
\ee
For the numerical analysis of (\ref{jgjjbnwb}), we use the more
convenient representation 
\be
S_2(y)= 2 \int_0^y (y-x) G(x) dx~ .    \label{14}
\ee
The question we address is whether and how $S_2(y)$ can be
approximated by the power law 
\be
S_2(y)= A_2 y^{\zeta(2)}~ .  \label{16}
\ee

\subsection{Properties associated with the second-order moment}

First, let us determine the conditions on $G(x)$ for which 
the form (\ref{16}) is exact. For arbitrary behaviors of $S_2(y)$,
one can always introduce a local exponent defined by
\be
\zeta(2,y)= {d \ln S(y) \over d \ln y}~ ,  \label{15}
\ee
which recovers $\zeta(2,y)=\zeta(2)$ if the power law (\ref{16}) holds
exactly. In the general case where $S_2(y)$ is not a pure power law,
we can write 
\be
\zeta(2,y)= {2 \over 2-\Delta(y)}~ ,   \label{17}
\ee
where
\be
\Delta(y)= {2 X(y) \over y}~, \qquad X(y)= {\displaystyle \int_0^y x
G(x) dx \over \displaystyle \int_0^y G(x) dx} ~ .   \label{18}
\ee
Thus, $\zeta(2,y)$ is independent of $y$ and the scaling law (\ref{16}) is exact
if $X(y)$ is proportional to $y$, where $X(y)$ has the interesting
interpretation of being the barycenter of the segment $[0,y]$ whose
mass density is $G(y)$. 

It is easy to show that $X(y) \sim y$ if and only if
$G(y)$ is a constant or a pure power law $G(y) \sim y^{-\lambda^2}$,
with some exponent that we denote $0< \lambda^2 <1$ for a reason that will
be clear soon.
In the former case, $X(y)=y/2$, $\Delta(y)=1$, which yields the
non-fractal scaling $\zeta(2,y)=2$. From (\ref{12}), we see that 
$G(y)$ is a constant if the covariance function $C(\tau)$ of the Gaussian process $\omega(t)$ 
defined in (\ref{nbjvnvqw}) is also a constant that is 
$dC(\tau)/d\tau =0$. Taking the derivative of 
(\ref{nbjvnvqw}) with respect to $\tau$, integrating by part and equating to 
zero for arbitrary values of $\tau$ imposes the condition $h(0)=0$, which 
just means that the measure $\mu(t)$ is uniform, hence the trivial
exponent $\zeta(2,y)=2$. The other case $G(y) \sim y^{-\lambda^2}$ is
more interesting. This yields
\be
X(y)= {1- \lambda^2 \over 2- \lambda^2}~ y ~ \rightarrow ~
\Delta(y)= 2~ {1- \lambda^2 \over 2- \lambda^2} ~ \rightarrow
~ \zeta(2,y)= 2-\lambda^2 ~ .
\ee
From (\ref{12}), we see that this case corresponds to $C(\tau)$ being an exact
logarithmic function of $\tau$, which is the property already mentioned
above with (\ref{7}) at the origin of the exact multifractality of the process $\delta_{\tau} X(t)$.

This discussion shows that, for $\varphi>0$, the scaling (\ref{16}) of the
second-order moment can only hold approximately at best.
In particular, the local exponent $\zeta(2,y)$ has a simple regular
behavior at the two boundaries $y\to 0$ and $y\to\infty$. Indeed, due to the limits
\be
\lim_{x\to 0} G(x)=1~ , ~~ \lim_{x\to \infty} G(x)= \langle\mu\rangle^2 / \langle\mu^2\rangle~ ,
\ee
we have
\be
S_2(y)\simeq y^2 \quad (y\to 0), \qquad S_2(y)\simeq {\langle\mu\rangle^2 \over \langle\mu^2\rangle} ~y^2
\quad (y\to\infty)~ ,
\ee
so that
\be
\lim_{y\to 0} \zeta(2,y)= \lim_{y\to \infty} \zeta(2,y)=2 ~ .
\label{nhmna;;q}
\ee
Borrowing the terminology from hydrodynamic turbulence, these two limits
(\ref{nhmna;;q}) implies the existence of an effective ``viscous'' scale
 $\tau_v$ and of an integral
scale $\tau_i$, such that if $\tau\lesssim \tau_v$ and $\tau\gtrsim
\tau_i$, then the increments (\ref{5}) are not multifractal. 
Note that the limit $y \to \infty$ is attained by taking 
$\ell \to 0$ at fixed $t$ since $y=t/\ell$. The limit $\ell \to 0$ thus
recovers a trivial regime, in contrast with the MRW in which the role
of $\ell$ is played by the scale of resolution which, when going to zero
yields a non-trivial genuine multifractal limit.

In the sequel,
we will be interested in the ``inertial range'' which exists if
$\zeta(2,y)$ is very slow function of the dimensionless variable
$y=\tau/\ell$ over the interval from some $y_0>1$ to some $y_1 \gg y_0 > 1$. In this 
case and if $\zeta(2,y)<2$, we will be entitled to speak about the multifractal
behavior of the random process $\delta X(t)$ for some wide range of scales and
define the effective exponent $\zeta(2)$ as the minimal value
\be
\zeta(2)= \min_{y} \zeta(2,y)~ ,
\label{20}
\ee
as we derive below as an optimal and efficient definition. Having obtained 
$\zeta(2)$, we can obtain a first
estimation of the intermittency coefficient $\lambda^2$
through the relation 
\be
\lambda^2= 2-\zeta(2)~ ,
\label{13}
\ee
which assumes that the parabolic dependence (\ref{8}) holds.

\subsection{Method of determination of the effective exponent $\zeta(2)$}

As we pointed out, the absence of
exact multifractility for $\varphi>0$ does not prevent the process from
exhibiting approximate multifractal scaling which, for all practical 
purposes, can be undistinguishable from an exact one. Indeed,
empirical data is always noisy and power law scaling are always sampled
on a finite (often small) range of scales \cite{Mandelfur,Biham,Biham2}.
The experimentally relevant question is thus to define scaling
from an operational view point consistent with what is done empirically.
We now describe a natural and robust determination of the approximate
scaling which will be seen to link intrinsically 
the definition and determination of the exponent $\zeta(2)$
with the existence of an integral time scale $L$ defined
from the range over which the approximate power law scaling holds.

For this, let us consider a $y$-interval $[y_1, y_2]$. We define $\zeta(2, y_1, y_2)$
over this interval as
\be
\zeta(2, y_1, y_2) = {\rm Arg}~\min_{\zeta} \int_{y_1}^{y_2} dy ||\ln S_2(y) - (A + \zeta  \ln(y)||^2~.
\label{A}
\ee
In words, expression (\ref{A}) determines $\zeta(2, y_1, y_2)$ as the exponent $\zeta$
which minimizes in a OLS sense the distance between $\ln S_2(y)$ and a straight
line in the $\ln y$-variable over the interval $[y_1, y_2]$. Let us now
introduce the parameter $\eta$, which measures the precision with which 
an approximate power law scaling is qualified. Specifically, a power law
with exponent $\zeta(2, y_1, y_2)$ is qualified 
if 
\be
\min_{\zeta} \int_{y_1}^{y_2} dy ||\ln S_2(y) - (A + \zeta  \ln(y)||^2 < \eta~.
\label{onvcoa}
\ee
Otherwise, the power law scaling is rejected. For a fixed $\eta$, we scan
all possible values of $y_1$ and $y_2$ for which 
condition (\ref{onvcoa}) is verified and we select the couple $(y_1, y_2)$ such that
the range $y_2/y_1$ is maximum. We thus obtain an approximate power law
scaling with apparent exponent $\zeta(2, y_1, y_2)$ within the confidence or noise
level $\eta$ over the maximum range $[y_1, y_2]$. 
We believe that this procedure embodies in a precise way the general fitting procedure
of experimental data. 

Let us study the limit $\eta \to 0$. In the case where $S_2(y)$ is not a pure
power law (and thus $\ln S_2(y)$ is not a perfect straight line $\ln y$), the condition
(\ref{onvcoa}) imposes that $y_2/y_1 \to 1$. Therefore, $\zeta(2, y_1, y_2 \to y_1)$
determined by (\ref{A}) yields the local slope of the function $\ln S_2(y)$
in the variable $\ln y$, in other words
\be
\zeta(2, y_1, y_2 \to y_1) = {d \ln S_2 \over d \ln(y)}|_{y_1}~.
\label{mgbnqrelb}
\ee
Consider now the function
\be
f_{y_1, y_2 \to y_1}(y_1)  =  \ln  S_2(y_1) -  \zeta(2, y_1, y_2 \to y_1)~ \ln(y_1)~.
\ee
This function $f_{y_1, y_2 \to y_1}(y_1)$ has at least one minimum in the
interval $y \in [\ell, +\infty]$, which we call $y_0$.
Close to its minimum, the function $f_{y_1, y_2 \to y_1}(y_1)$  
can be expanded up to second order to obtain
\be
f_{y_1, y_2 \to y_1}(y_1)   =  A +  (1/2) a ( \ln(y_1) - \ln (y_0))^2~,
\ee
where 
\be
a = d^2 \ln S_2 / d^2 \ln(y) |_{y_0}
\label{mnhnn;}
\ee
is the second-order derivative of $\ln S_2$
with respect to $\ln y$, estimated at $y_0$. It is clear that, 
for small finite values of $\eta$, the largest range
$y_2/y_1$ is obtained for $y_1=y_0$ (we assume here that $f$ is convex
so that the local minimum is the global minimum; for the range
of $y$'s in the examples we have investigated, we have found the convexity
condition to always hold). In addition, 
$f_{y_1, y_2 \to y_1}(y_1)$ does not change appreciably over a range of 
$\ln(y_1)$ proportional to $1/a^{1/2}$. Thus, the smallest $a$ is, the largest 
is the range over which $f_{y_1, y_2 \to y_1}(y_1)$ will be almost constant 
and thus over which the exponent $\zeta(2)$ will well-defined and constant. This reasoning
provides the algorithm to measure the approximate exponent $\zeta(2)$
for a given data set, which we are going to use in a systematic way. 
It is given by (\ref{mgbnqrelb}), where
$y_1$ is chosen equal to the argument $y_0$ such that the second derivative (\ref{mnhnn;}) 
is zero (or reaches the minimum positive value over the whole range available when
zero is not crossed). This is equivalent to searching for the exponent (\ref{mgbnqrelb})
which takes the smallest possible positive value over the range of study.

\section{Scaling of the second-order moment $S_2$}

We study the process (\ref{5}), whose properties are controlled by the two
key parameters $\varphi$ defined in (\ref{10}) and $\sigma^2$ defined in 
(\ref{11}). We calculate the second-order structure function $S_2(y)$
defined in (\ref{14}) with $y=\tau/\ell$, where $G(x)$ is obtained from expressions (\ref{12})
and (\ref{nbjvnvqw}).

As a first example, we fix $\varphi=0.01$ and scan $\sigma^2=20;30;40$.
The corresponding structure functions $S_2(y)$ are plotted in figure 1
as a function of $y=\tau/\ell$ in log-log scales, together with the best
power law fits shown as straight lines. A superficial examination
suggests an excellent scaling behavior over at least four orders of
magnitudes, with an exponent which clearly varies with $\sigma^2$ from
$1.66$ to $1.34$ when $\sigma^2$ goes from $20$ to $40$. This property
is novel compared with the MRW and previous multifractal process with
$\varphi=0$ and, as we are going to explore in some length, allows us to
control the multifractal properties continuously as a function of
$\sigma^2$ as well as $\varphi$. Fig. 2 examines more precisely the
nature of the apparent power law behavior depicted in Fig. 1 by plotting
the local exponent $\zeta(2,y)$ defined in (\ref{15}) as a function of
$y$ in log-scale in the range $y \in [1,10^7]$, for a fixed
$\sigma^2=10$ and varying values of $\varphi = 0.002$ to $0.01$. The
first important message of this figure is that the exponent $\zeta(2,y)$
is approximately constant over a large range of $y$-values, all the more
so, the closer $\varphi$ is to zero (this later property is of course
not a surprise since $\varphi \to 0$ recovers previously known
multifractal processes). Interestingly, this approximately constant
value for $\zeta(2,y)$ has a rather large dependence on $\varphi$
itself, showing again that we can control the intermittency parameter by
changing $\varphi$ as well as $\sigma^2$. Another important observation
is the sharp variation of $\zeta(2,y)$ on the left-hand-side of the range,
suggesting a rather well-defined ``viscous scale'' $\tau_v$, which we
characterize as the boundary between the region of approximate constancy
of $\zeta(2,y)$ around the definition (\ref{20}) and the region of sharp 
increase of $\zeta(2,y)$ as $y$ decreases below the minimum (\ref{20}).
In the examples shown in fig. 1, $\tau_v$ is in the range $10^2\div
10^3$. The closer $\varphi$ is to $0$, the smaller is the viscous scale 
$\tau_v$ and the better is the multifractal scaling.
In contrast, it is not obvious to identify the integral scale
over the interval shown in Fig. 1 as the increase of $\zeta(2,y)$ towards $2$
has not yet occurred appreciably even up to $y=10^7$, ensuring a rather nice
approximate scaling with anomalous multifractal exponent $\zeta(2)<2$. 
Another way to express this observation is that the ``inertial regime'' 
over which the scaling of the second-order moment holds has not absolutely
defined boundaries. Therefore, the transition to the integral scale is smooth,
a property also documented in hydrodynamic turbulence \cite{Frisch}.

Fig.~3 is the same as Fig.~1 but for a large value of $\varphi=0.5$
and with $\sigma^2=1$ and $5$. It shows that the scaling law for
$S_2$ still holds over approximately two decades in the horizontal $y$ scales
(and of course more in the vertical scale). A plot analogous to Fig.2 for
these values of $\varphi$ and $\sigma^2$ shows obviously larger deviations
from a constant behavior (not shown). As we discussed in the previous
section, the important message is that we have a rather large latitude
in changing $\varphi$ and $\sigma^2$ to exhibit a reasonable (with 
respect to standard experimental precision) multifractal scaling.

Fig.~4 shows the other limit of a small value of $\varphi=0.001$ for
a large range of values of $\sigma^2=100$ to $500$. While we expect
indeed that the multifractal scaling should extend on large
ranges as $\varphi$ decreases to zero, the most remarkable fact is
that the exponent $\zeta(2)$ can be continuously adjusted at will from 
$2$ all the way to close to $1$ by varying $\sigma^2$ at fixed $\varphi$.
Note also the large range of $y$-scale over which the apparent exponent 
$\zeta(2,y)$ remains approximately constant: for instance, for $\sigma^2=100,
\varphi=0.001$, we measure a well-defined constant exponent $\zeta(2)=1.81$ over
$8$ orders of magnitudes. Even for these very large ranges of scales, the
known existence of an integral scale and the transition to the normal 
value $\zeta(2)=2$ is not seen. 

Fig.~5 gives a synopsis of the dependence of the effective exponent $\zeta(2)$
as a function of the two control parameter $\varphi$ and $\sigma^2$. Actually,
we show instead the ``intermittent coefficient'' $\lambda^2= 2-\zeta(2)$
first introduced in equation (\ref{13}), in analogy with the parabolic 
multifractal spectrum (\ref{8}). The most important observation is that
$\lambda^2$ increases linearly with $\sigma^2$ before saturating to $1$ asymptotically
for large $\sigma^2$'s. The upperbound $1$ for $\lambda^2$ results
from the following property of the second-order moment $S_2$ obtained
from the definition (\ref{14}):
\be
{d^2 S_2(y) \over dy^2}= 2 G(y)>0~ .
\ee
The fact that the second-order derivative of $S_2(y)$ is positive means that
$S_2(y)$ increases faster than a linear function of $y$, hence 
\be
\zeta(2)> 1 \qquad \rightarrow \qquad \lambda^2<1~ .  \label{22}
\ee

\section{Higher-order moments, universal scaling function and multifractal spectra}

\subsection{Definition and determination of the higher-order moments}

We have also investigated the higher-order moments $S_3$, $S_4$ and $S_5$
up to order $5$ of the increments $\delta_\tau X(t)$ defined in (\ref{5}).
Moments of arbitrary orders can be obtained from formulas generalizing 
expression (\ref{14}) for $S_2$, as follows
\be
S_q(y)= q(q-1) \int_0^y (y-x) G_q(x) dx ~ ,  \label{23}
\ee
where, for $q>2$,
$$
G_q(x)= G(x) \int_0^x du_1 \dots \int_0^x du_{q-2} 
$$
\be
\prod_{i=1 \atop
j=i+1}^{q-2} G(x_i) G(u-x_i) G(x_i-x_j)~.   \label{23bis}
\ee
The corresponding local exponents for the higher-order moments,
are defined analogously to (\ref{15}) and (\ref{17}) as
\be
\zeta(q,y)= {2 \over 2-\Delta_q(y)}~ ,
\ee
where
\be
\Delta_q(y)= {2 X_q(y) \over y}~ , \qquad X_q(y)= {\displaystyle
\int_0^y x G_q(x) dx \over \displaystyle \int_0^y G_q(x) dx} ~ .
\ee
Using these relations, we adopt the definitions generalizing (\ref{20})
for the effective exponents
\be
\zeta(q)= \min_{y} \zeta(q,y)    \label{24}
\ee
associated with the effective power laws
\be
S_q(y)= A_q y^{\zeta(q)}~ , \qquad A_q= S_q(y_{m}) y_m^{-\zeta(q)}~, \label{25}
\ee
where $y_{m}$ is the value of $y$ which makes $\zeta(q,y)$ minimum.

\subsection{Universal scaling ansatz}

Our numerical calculations of the higher-order moments ($q\leqslant 5$) show
that excellent scaling is observed for these moments over a wide range
of the dimensionless variable $y$, similarly to the case of the 
second-order moment $S_2$ presented in figures 1-3 \cite{citation}. 
This allows us to determine the dependence of 
the effective multifractal spectrum
$\zeta(q)$ defined by (\ref{24}) with respect to $q$,
as well as $\sigma^2$ and $\varphi$. For this, it is convenient to use
the parabolic spectrum (\ref{8}) as a proxy to extract an effective
(a priori) $q$-dependent intermittent coefficient defined by
\be
\lambda^2(q;\sigma^2,\varphi)=2 {q-\zeta(q) \over q(q-1)}~ .  \label{26}
\ee
We also make explicit in the notation $\lambda^2(q;\sigma^2,\varphi)$ the
dependence on the two parameters $\sigma^2$ and $\varphi$.
Similarly to the case of the second-order moment, expression (\ref{23})
allows us to show that $\zeta(q)$ and
$\lambda^2(q;\sigma^2,\varphi)$ satisfy to the following inequalities
\be
\zeta(q)>1 \qquad \rightarrow \qquad \lambda^2(q;\sigma^2,\varphi)<
2/q~ ,    \label{27}
\ee
which generalizes (\ref{22}) for arbitrary $q$'s.

Our detailed numerical calculations suggest the 
conjecture that the dependence of 
$\lambda^2(q;\sigma^2,\varphi)$ with respect to $\sigma^2,\varphi$
can be factorized as follows:
$\lambda^2(q;\sigma^2,\varphi)= 2\Lambda(aq)/q$, where the factor
$a=a(\sigma^2, \varphi)$ depends only on the
parameters $\sigma^2$ and $\varphi$ and not on $q$ and the function
$\Lambda(x)$ is a monotonously increasing. Moreover, the
numerical calculations of the intermittent coefficients show that 
it is an excellent approximation to represent $a(\sigma^2, \varphi)$
as a linear function of $\sigma^2$, i.e., $a= b(\varphi) \sigma^2$. 
This provides the following 
universal scaling law for the generalized intermittency coefficients 
\be
\lambda^2(q;\sigma^2,\varphi)={2 \over q} \Lambda(b \sigma^2 q)~ .
\label{28}
\ee
The factor $b$, which is independent of $q$ and of $\sigma^2$ according
to the scaling ansatz (\ref{28}), can be determined from the
intermittency coefficient $\lambda^2(q=2;\sigma^2,\varphi)=\Lambda(2b \sigma^2)$ for $q=2$,
previously reported. Since $\lambda^2(q=2;\sigma^2,\varphi)$ is a linear
function of $\sigma^2$ for not too large $\sigma^2$'s as shown in Fig.~5, this
implies that the function $\Lambda(x)\sim x$ is linear for small $x$'s.
Defining $b$ such that, for $x\to 0$, we have $\Lambda(x)\simeq x$, we obtain
\be
b(\varphi) \simeq {\lambda^2(q=2;\sigma^2,\varphi) \over 2 \sigma^2} \qquad (\lambda^2\ll 1)~.
\label{29}
\ee
Taking for instance $\sigma^2=20$ for which all intermittency coefficients
remain small, we find that $b(\varphi)$ is very well represented by the 
following power law dependence 
\be
b \simeq \alpha \varphi^\beta~ , \qquad \alpha= 0.58~ , \quad \beta=0.92~ .
\label{mgbmlqqasfgg}
\ee
Fig.~6 plots the reconstructed
functions $\Lambda(x)=(q/2)\lambda^2(q;\sigma^2,\varphi)$ for different orders $q$.
The excellent collapse provides a first validation of the scaling ansatz 
(\ref{28}).

\subsection{Multifractal spectra}

The general scaling ansatz (\ref{28}) allows us to express
the multifractal spectrum $\zeta(q)$ under the following form
\be
\zeta(q)= q + (1-q) \Lambda(b \sigma^2 q)~ .  \label{30}
\ee
Moreover if, for some particular multifractal phenomenon,
the intermittency coefficient $\lambda^2=\lambda^2(2)$ is small
compared to $1$, then one
can rewrite expression (\ref{30}) in the more universal form
\be
\zeta(q)=q+ (1-q) \Lambda(\lambda^2 q/2) ~,  \label{31}
\ee
whose dependence on  $\sigma^2$ and $\varphi$ is completely
embedded in that of $\lambda^2$. This implies that the knowledge of
the intermittency coefficient $\lambda^2$ together with the scaling
function $\Lambda(x)$ allows one to determine the multifractal 
spectrum for arbitrary orders even for $q\gg 1$. 

Our prediction (\ref{31})
can be checked by direct numerical calculations of the moments 
up to order $5$ that we have performed.
Fig.~7 plots the multifractal spectra $\zeta(q)$ for $q=0$ to $5$
obtained by two methods: (i) the circles are the direct numerical 
integration of (\ref{23},\ref{23bis}) of the stochastic process
for $\varphi=0.004$ and different values of $\sigma^2$; (ii) the continuous lines are 
obtained by using (\ref{30}) with the scaling function $\Lambda(x)$
constructed as in Fig.~6 for $q=2$ and $\varphi=0.001$ and applied to the
case $\varphi=0.004$. The agreement is good, even for the large values $q=5$.
This illustrates that different combinations of $\varphi$ and $\sigma^2$
with fixed $b \sigma^2$ give the same value of the intermittency coefficient
$\lambda^2$ (for $\lambda^2$ not too large) according to (\ref{28})
and thus the same multifractal spectrum $\zeta(q)$.

Note that $\zeta(q)$ becomes non-concave for large $\sigma^2$'s at large $q$'s, a
property which is excluded for exact multifractal scaling by the H\"older inequality
applied to the moments  $\langle [\delta_\tau X(t)]^q \rangle$. The absence
of concavity in a certain range of parameters reflect the fact that 
multifractality is not an asymptotic property observed at small or large values
of the dimensionless variable $y$, but only in some intermediate scaling range.
Non-concavity also prevents obtaining the exact multifractal 
spectrum of dimensions $f(\alpha)$ of singularities $\alpha$, but yields
only a concave envelop of it \cite{touchette}.

\section{Numerical and Mathematical insights on the origin of the effective
multifractality}

As we showed, these multifractal properties are only effective
properties or approximations, as exact multifractality only holds when
$C(\tau)$ defined in (\ref{nbjvnvqw}) is proportional to the logarithm of $\tau$
(up to an integral scale).
The approximate multifractality discussed here can be tracked back to
the approximate logarithmic dependence of $C(\tau)$, as we know make clear,
numerically and mathematically.

Let us consider the second-order moment $S_2(y)$. The scaling (\ref{16})
is a precise description of $S_2(y)$ if $G(y)$ in (\ref{12}) is close to a power law, 
i.e., $\sigma^2 d(y)$ is close to $\lambda^2 \ln y$. To test if this is the case
for different values of $\sigma^2$ and $\varphi$, we construct 
\be
\delta(y)= \sigma^2 d(y)/\lambda^2  ~ ,  \label{c}
\ee
which should be close to $\ln y$ to justify our results above.
Recall that, for rather small $\sigma^2$ (actually, we just need
that $\sigma^2 \lesssim 100$), we have the relation
(\ref{29}) with (\ref{mgbmlqqasfgg}), which together with (\ref{c})
yields 
\be
\delta(y)= {d(y) \over b(\varphi)}= {1-C(y) \over  b(\varphi)} ~ ,
\label{d}
\ee
which has to be almost equal to $\ln y$ for any $\sigma^2$ and $\varphi$ to
justify our results. This is verified in Fig.~8,
which plots $\delta(y)$ given by (\ref{d}). As $\varphi$ departs more
and more from $0$, increasing
deviations from $\ln y$ occur, which reduce the range over which
scaling and multifractality can be observed. An argument similar
to that presented in Fig.~8 was proposed in \cite{PRL05,JGR05}
for the multi-scaling of the conditional response function (Omori law).

This argument can be made mathematically precise as follows. 
Let us consider the function 
\be
K(y,\varphi)= \int_0^\infty h(x,\varphi) h(x+y,\varphi) d x~ , 
\label{eqq1}
\ee
where we re-define
\be
h(x,\varphi)= { 1 \over (1+x)^{\varphi+1/2}} ~,  \label{eqq2}
\ee
such that
\be
\sigma^2 \equiv K(0,\varphi)= {1 \over 2 \varphi}~ .  \label{eqq3}
\ee
Determining how well the covariance function $C(t)$, defined by (\ref{nbjvnvqw}),
is approximated by a logarithm of $t$ amounts to study how well 
$K(y,\varphi)$ is approximated by a logarithmic function of $y$.

The study of $K(y,\varphi)$ is based on its explicit analytical expression
\be
\begin{array}{c}
K(y,\varphi)= 
\displaystyle {\sqrt{\pi}~ \Gamma(\varphi) \sec(\pi \varphi) \over 2
\Gamma(\varphi+1/2)} \left(- {2\over y}\right)^{2 \varphi} \\
- 2 {(1+ y)^{-\varphi+1/2} \over 1-2 \varphi} F\left(1, { 1 \over 2}+ \varphi,  { 3
\over 2}- \varphi, 1+y\right)~ ,
\end{array}
\label{mgkmfvv;q}
\ee
where $F(a,b,c,z)$ is a hypergeometric function.

Let us introduce the analog of the function $d(y)$ (see (\ref{12})) defined by
\be
D(y,\varphi)=\sigma^2- K(y,\varphi)= {1\over 2\varphi} - K(y,\varphi)~. 
\ee
Using (\ref{mgkmfvv;q}), we obtain the following asymptotic expression for $D(y,\varphi)$:
\be
D(y,\varphi)\simeq  -{1 \over 2} {F(y,\varphi)-F(y,0) \over \varphi} + \ln
\left( {\sqrt{1+y}+1 \over \sqrt{1+y}-1}\right)~ ,
\label{eqq4}
\ee
valid for $\varphi\to 0$,
where
\be
F(y,\varphi)= \rho(\varphi)  \left({2 \over y} \right)^{2\varphi}~ ,
\label{eqq5a}
\ee
and
\be
\rho(\varphi) = \varphi \Gamma(\varphi) {\sqrt{\pi} \over \Gamma(1/2+\varphi)}~.
\sec(\pi\varphi)~ . 
\label{eqq5b}
\ee
We verify that the asymptotic expression (\ref{eqq4}) gives an excellent
approximation to the exact expression for $D(y,\varphi)$ obtained by using
(\ref{mgkmfvv;q}) for $1 \leq y \leq 10^{15}$ for $\varphi \lesssim 0.1$,

To reveal the dependence of $D(y,\varphi)$ with respect to $y$ for small
$\varphi$, we may interpret the first fraction in the r.h.s. of
the relation (\ref{eqq4}) as a discrete approximation of the derivative 
$F'(y,\varphi)$ of the function $F(y,\varphi)$ with respect to
$\varphi$. This leads to the following approximate relation
\be
D(y,\varphi) \simeq -{1 \over 2} F'\left(y,{\varphi \over 2} \right)+ \ln
\left( {\sqrt{1+y}+1 \over \sqrt{1+y}-1}\right) ~ ,
\label{eqq6}
\ee
where 
\be
F'(y,\varphi)= -2 \ln\left[A(\varphi) y^{\rho(\varphi)} \right] \left({2
\over y} \right)^{2\varphi} ~ ,
\ee
with
\be
A(\varphi)= \left({1 \over 2}\right)^{\rho(\varphi)} \exp\left(-
{\rho'(\varphi) \over 2}\right)~ .
\ee

Since $\ln \left[\left(\sqrt{1+y}+1 \right) / \left( \sqrt{1+y}-1\right)\right] =
2/\sqrt{y}+ {\cal O}(1/y)$ for $y \gg 1$, we can neglect this logarithmic term in 
(\ref{eqq6}) for $\varphi < 1/2$ and obtain
the following logarithmic-power approximation 
\be
D(y,\varphi) \simeq \ln\left[A(\varphi/2) y^{\rho(\varphi/2)} \right]
\left({2 \over y} \right)^{\varphi}~ .  \label{eqq7}
\ee
Figure 9 plots the exact function $D(y,\varphi)$ and its approximation (\ref{eqq7}),
showing the approximate linear dependence of $D(y,\varphi)$ as a function of $\ln y$
for a large range of $y$ for $\varphi>0$.

For $\varphi \to 0$, $\rho(\varphi/2) \to 1$ and 
we recover $D(y,\varphi \to 0)  \to \ln y$, as expected.
For $\varphi>0$, expression (\ref{eqq7}) makes explicit
the deviation from a pure logarithmic dependence of $D(y,\varphi)$
in the form of the power factor $\left(2 / y \right)^{\varphi}$.
The predicted deviation from a pure logarithm gives us the possibility
to estimate the integral scale beyond which the effective multifractality
breaks down. We thus define the
integral scale $L_{\epsilon}$ as the solution of the equation
\be
\left({2 \over y} \right)^\varphi = \epsilon~,
\ee 
where $\epsilon < 1$ measures the deviation of $D(y,\varphi)$ from
$\ln y$. This gives $L_{\epsilon} = 2/\epsilon^{1/\varphi}$.
Taking arbitrarily $\epsilon=1/2$, we obtain $L_{1/2} \sim 2000$
for $\varphi=0.1$, $L_{1/2} \sim 1.3 \times 10^{30}$ for 
$\varphi=0.01$ and the astronomical value
$L_{1/2} \sim 10^{301}$ for $\varphi=0.001$. These values
explain why the saturation of the effective multifractal scaling 
predicted for large $y$ is not observed for small $\varphi$'s.

\section{Concluding remarks}

We have confirmed on the multifractal spectrum $\zeta(q)$ the
proposition previously introduced in the context of a model of earthquake
\cite{PRL05,JGR05} that processes constructed as exponentials of
long-memory processes should exhibit multifractal properties over
a significant range of the parameters. While the
initial argument concerned the time decay of the conditional expectation
of the response function (called the Omori law for the decay of the
number of aftershocks in the context of seismology) as a 
generalization to the prediction and observation for the multifractal
random walk discussed in Ref.~\cite{SMM}, we have extended
the analysis to show that multifractality is a robust property 
of this class of processes defined as exponentials
of long memory processes. 

Notwithstanding the fact that the multifractal properties discusses here
are only effective properties or approximations, we claim that there is
probably no way to tell the difference between an exact multifractal
random walk or an exact multifractal scaling from our approximate 
scaling laws and approximate $\zeta(q)$ functions, given the noise 
and the limited ranges usually seen
in experimental and numerical studies. We have illustrated this claim
by showing that a given multifractal spectrum $\zeta(q)$ can be obtained
for several sets of parameters of our model.
In a sequel, we will discuss what observables should be used
to possibly distinguish between different implementations, so as to 
extract useful physical interpretation of the parameters $\lambda^2$,
$\varphi$ and $\sigma^2$.

{\bf Acknowledgments:} We acknowledge stimulating discussion with
J.-F. Muzy and G. Ouillon and thank J.-F. Muzy for a critical
reading of the manuscript. This work is partially supported
by NSF-EAR02-30429, and by the Southern California
Earthquake Center (SCEC). SCEC is funded by NSF
Cooperative Agreement EAR-0106924 and USGS Cooperative
Agreement 02HQAG0008. The SCEC contribution number for
this paper is XXX.

{}

\clearpage

\begin{quote}
\centerline{ \resizebox{14cm}{!}{\includegraphics{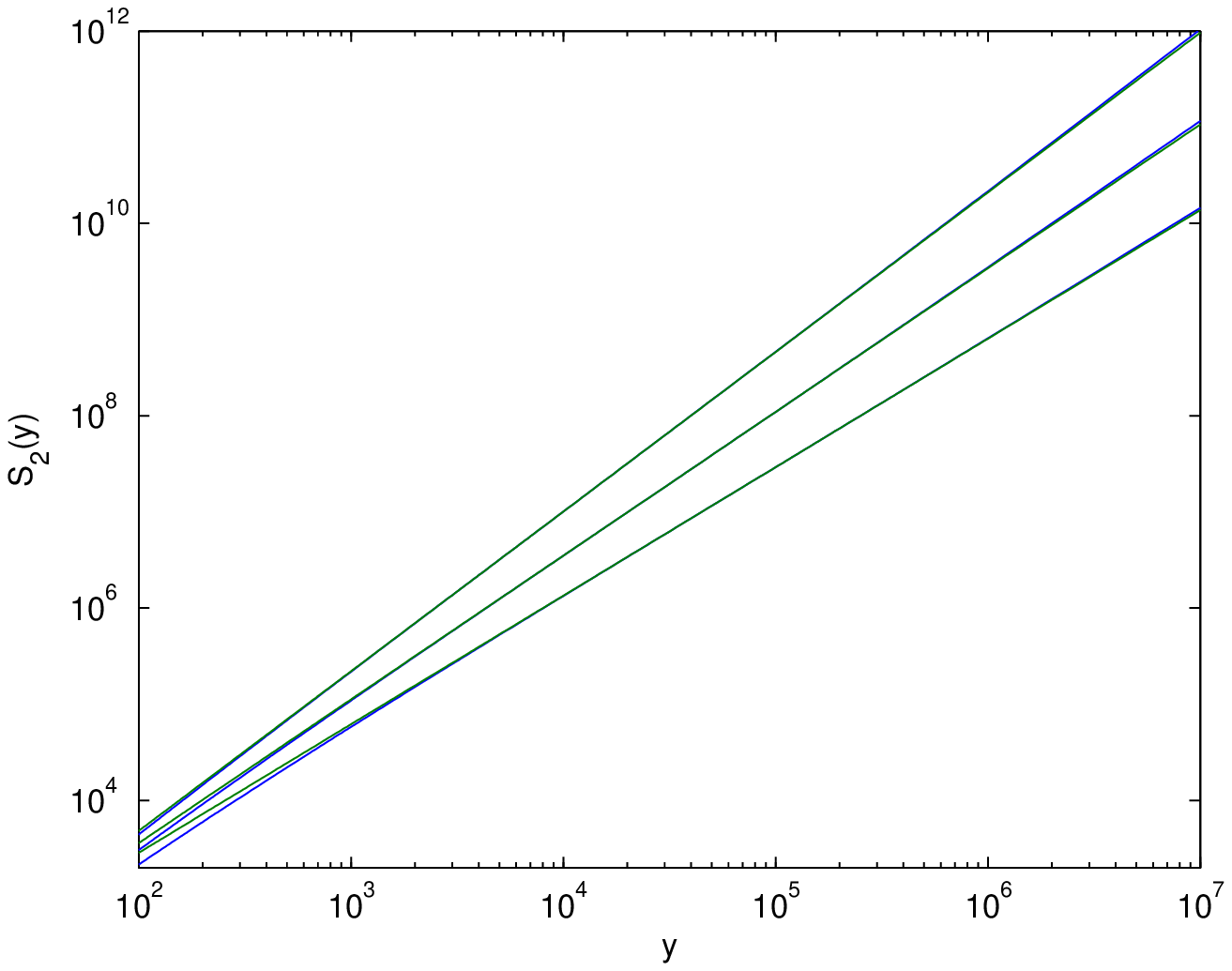}}} 
{\bf Fig.~1:} \small{Log-log-plot of the second-order moment $S_2(y)$ (\ref{14}) and its
power approximation (\ref{16}) for $\varphi=0.01$ and
$\sigma^2=20;30;40$ (top to bottom). The corresponding exponents are equal to
$\zeta(2)=1.66;1.49; 1.34$.}
\end{quote}

\clearpage

\begin{quote}
\centerline{ \resizebox{14cm}{!}{\includegraphics{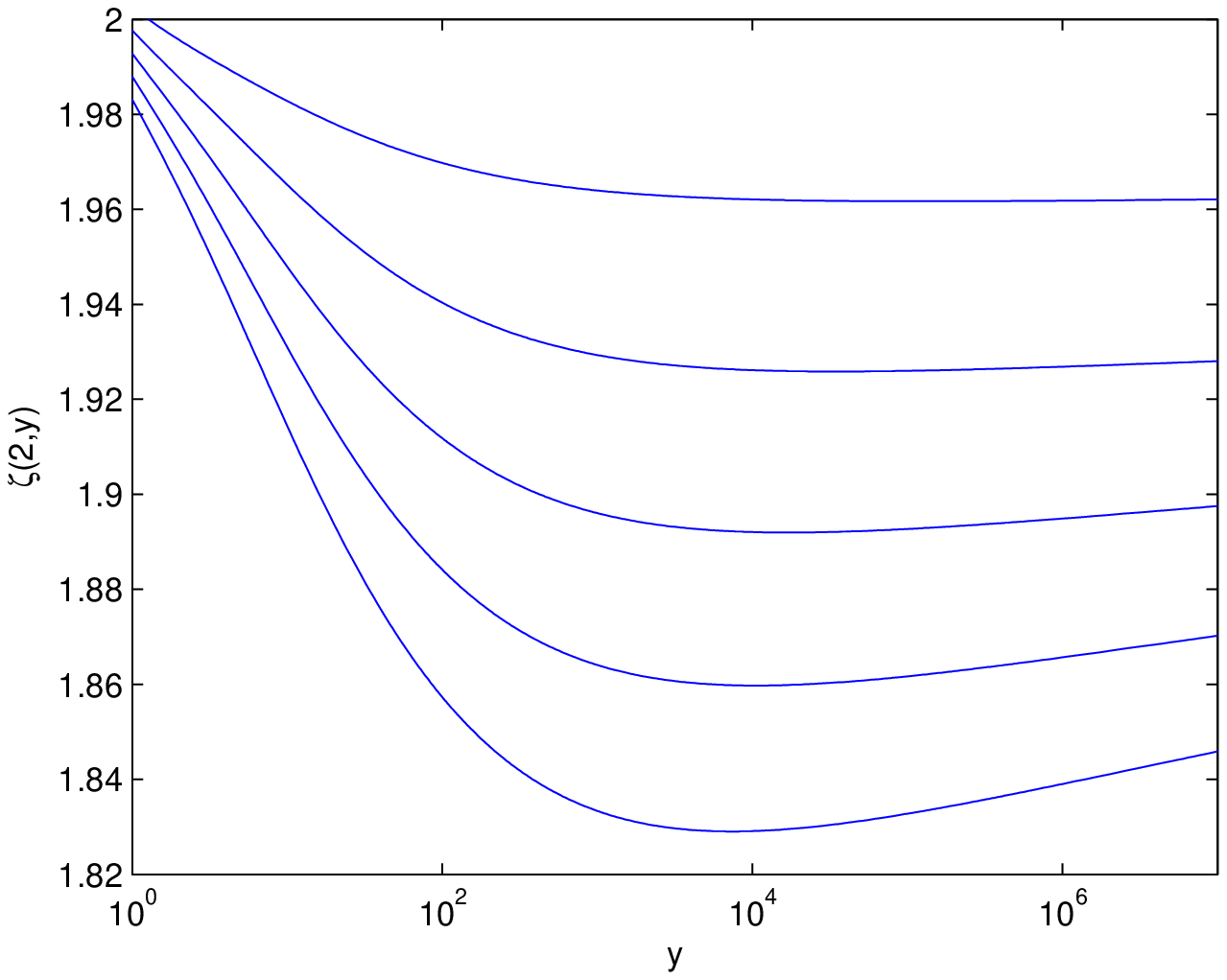}}} 
{\bf Fig.~2:} \small{Local exponent $\zeta(2,y)$ as a function of $y=\tau/\ell$
for $\sigma^2=10$ and $\varphi=0.002; 0.004; 0.006; 0.008; 0.01$ (top to bottom).}
\end{quote}

\clearpage

\begin{quote}
\centerline{ \resizebox{14cm}{!}{\includegraphics{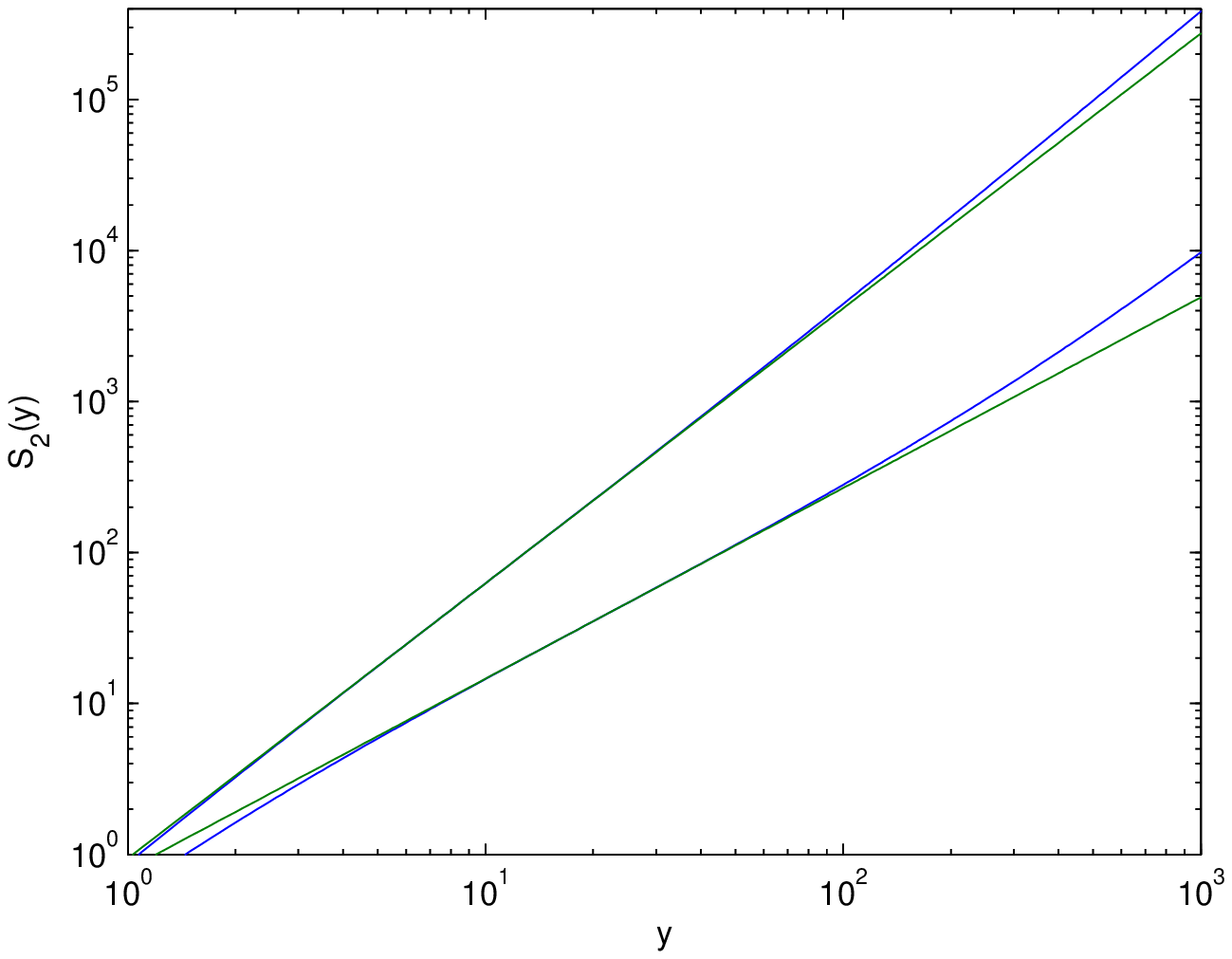}}} 
{\bf Fig.~3:} \small{Log-log plot of second-order moment $S_2(y)$ calculated
using (\ref{14}) and its
power approximation (\ref{16}) for $\varphi=0.5$ and $\sigma^2=1;5$ (top to bottom).
The corresponding effective exponents are respectively 
$\zeta(2)=1.82;1.26$.}
\end{quote}

\clearpage

\begin{quote}
\centerline{ \resizebox{11cm}{!}{\includegraphics{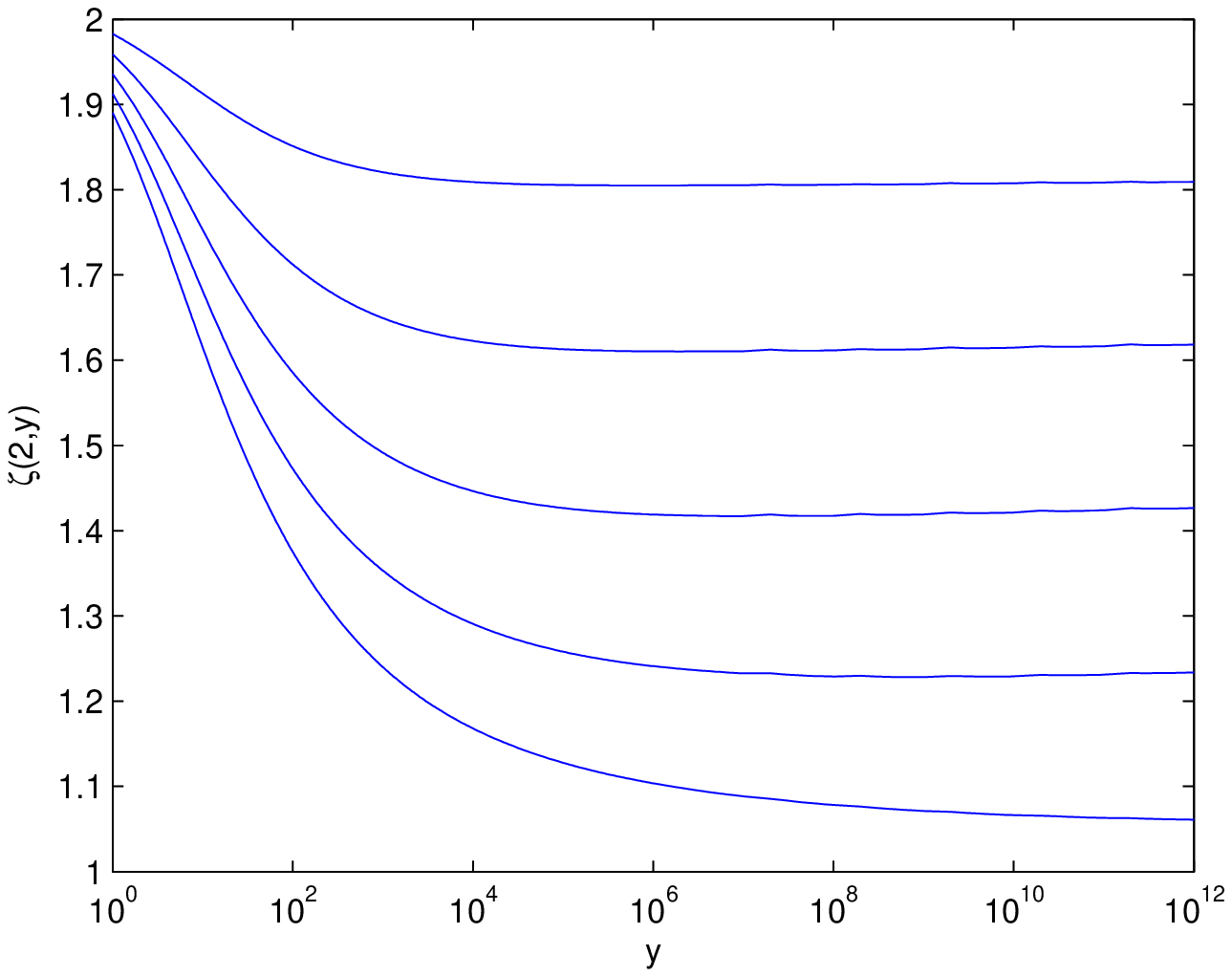}}} 
{\bf Fig.~4:} \small{Dependence of the local exponent $\zeta(2,y)$ 
as a function of $y=\tau/\ell$ in log-scale, for
$\varphi=0.001$ and $\sigma^2=100; 200; 300; 400; 500$ (top to bottom).}
\end{quote}

\clearpage

\begin{quote}
\centerline{ \resizebox{11cm}{!}{\includegraphics{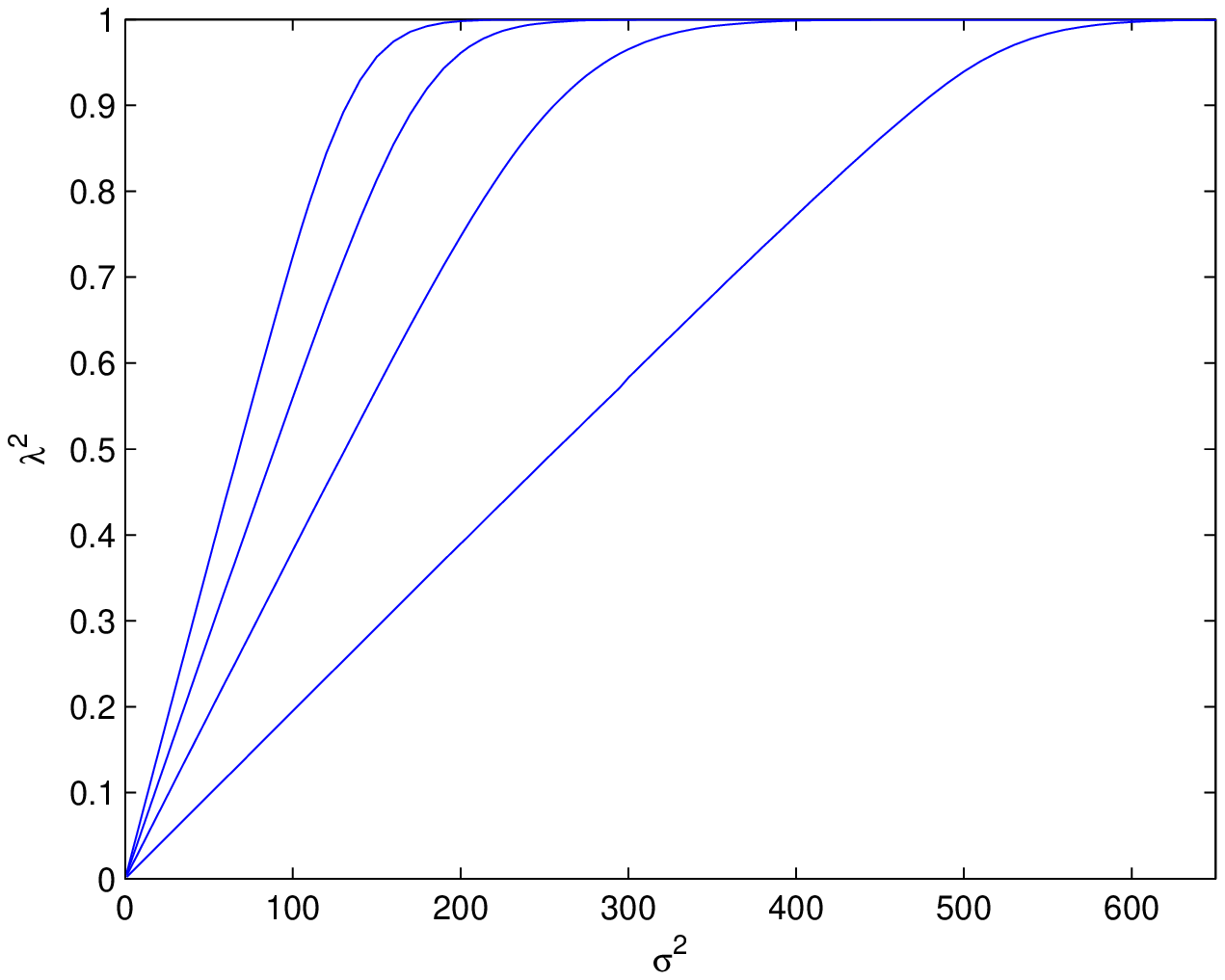}}} 
{\bf Fig.~5:} \small{Dependance of the intermittency coefficient
$\lambda^2 = 2-\zeta(2)$ as a function of $\sigma^2$
for different values of $\varphi=0.01\div 0.04$ (bottom to top).}
\end{quote}

\clearpage 

\begin{quote}
\centerline{ \resizebox{11cm}{!}{\includegraphics{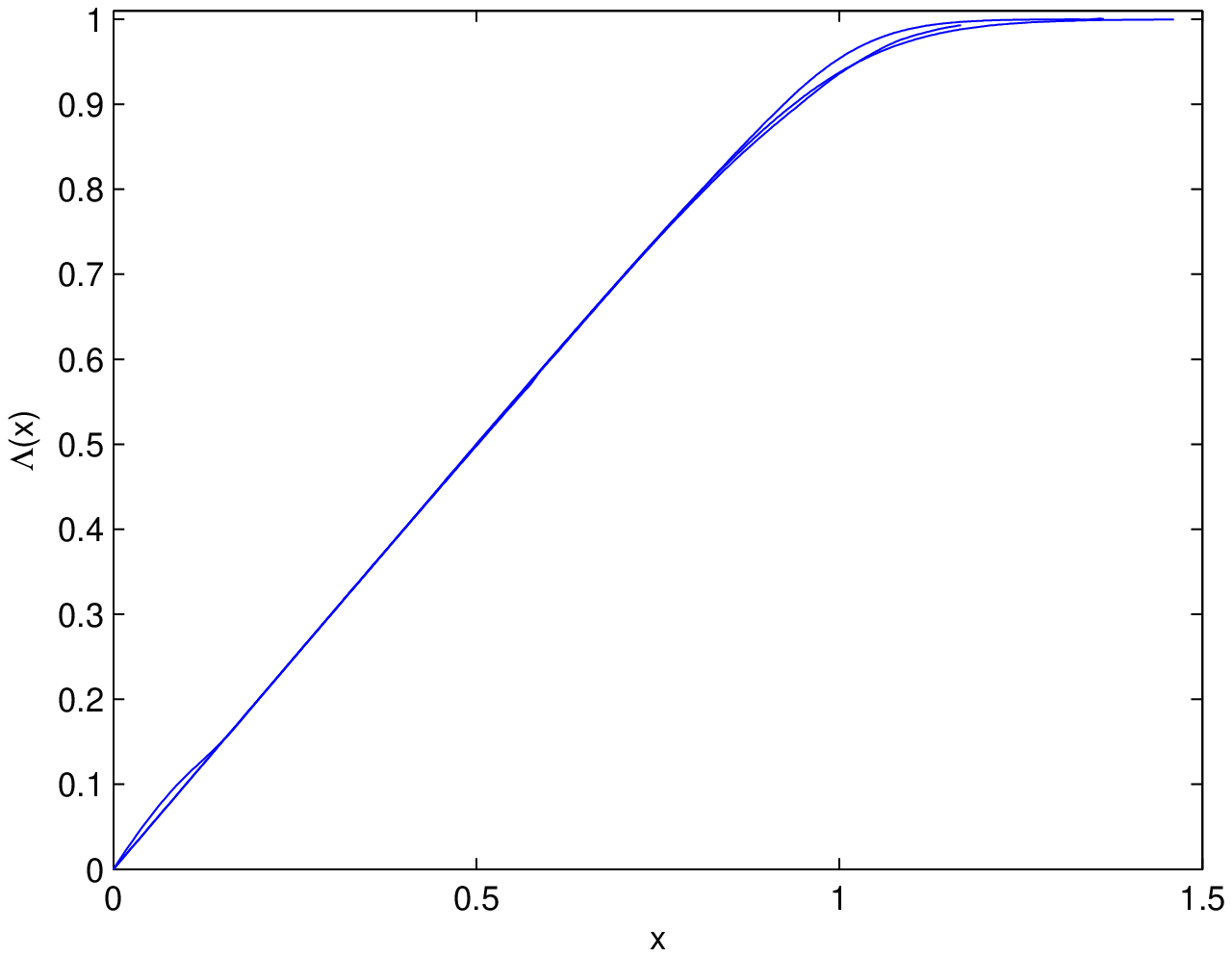}}} {\bf
Fig.~6:} \small{Plot of the universal scaling function $\Lambda(x)$, obtained from
relations (\ref{26}) and (\ref{28}) and the numerical calculation of the
dependence of the
effective multifractal exponents $\zeta(q)$ as a function of $\sigma^2$, for
$\varphi=0.001$ and $q=2;3;4$. The slight discrepancies between the curves in the 
neighborhood of $x=1$ can be attributed to some systematic errors of numerical calculations.
}
\end{quote}

\clearpage

\begin{quote}
\centerline{ \resizebox{12cm}{!}{\includegraphics{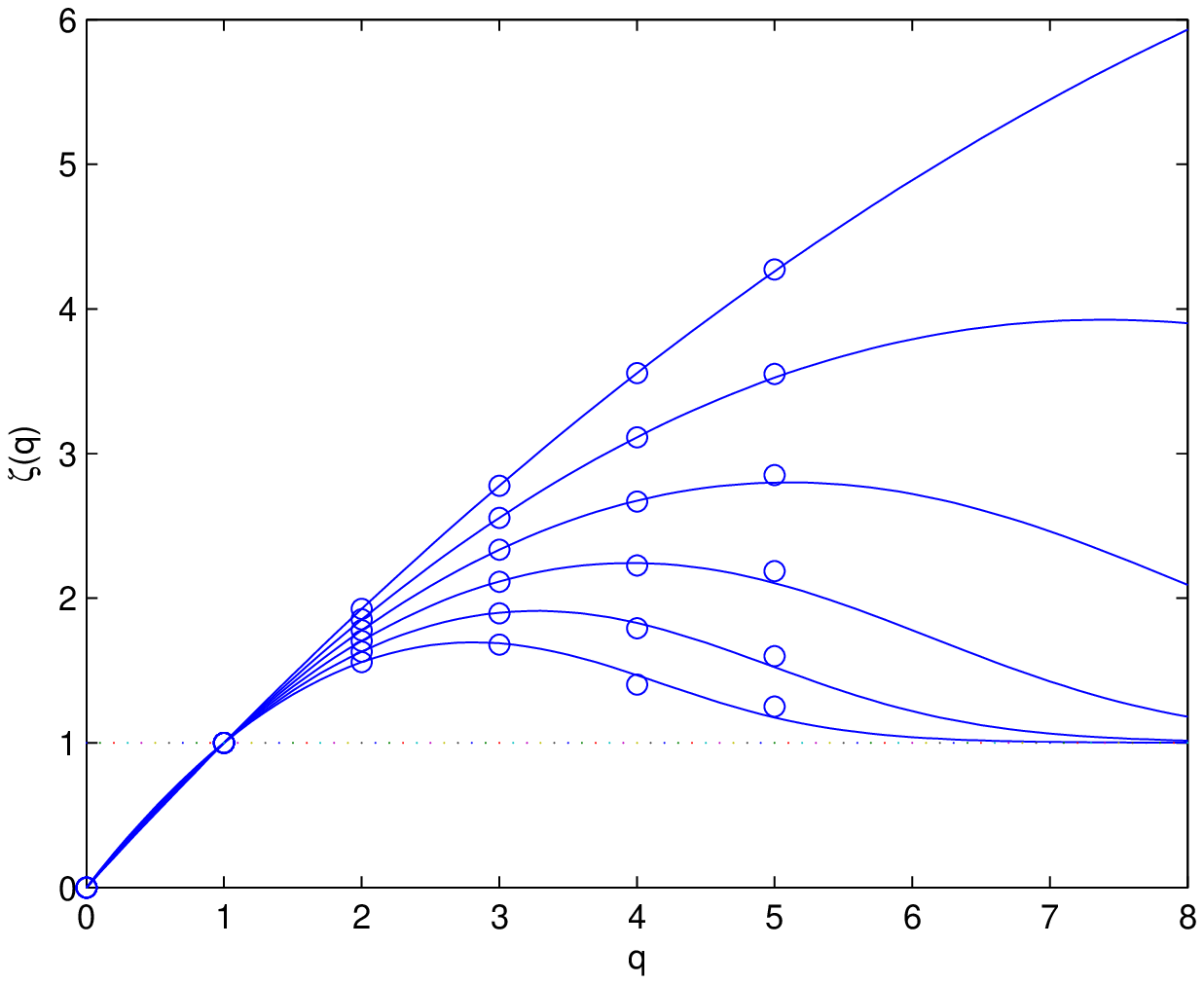}}} {\bf
Fig.~7:} \small{Universal multifractal spectra $\zeta(q)$ for 
$\varphi=0.004$ and $\sigma^2=10;20;30;40;50;60$ (top to bottom).
We compare two different methods for estimating $\zeta(q)$:
(i) the circles are the direct numerical 
integration of (\ref{23},\ref{23bis});
(ii) the continuous lines are 
obtained by using (\ref{30}) with the scaling function $\Lambda(x)$
constructed as in Fig.~6 for $q=2$.
}
\end{quote}

\clearpage

\begin{quote}
\centerline{\resizebox{14cm}{!}{\includegraphics{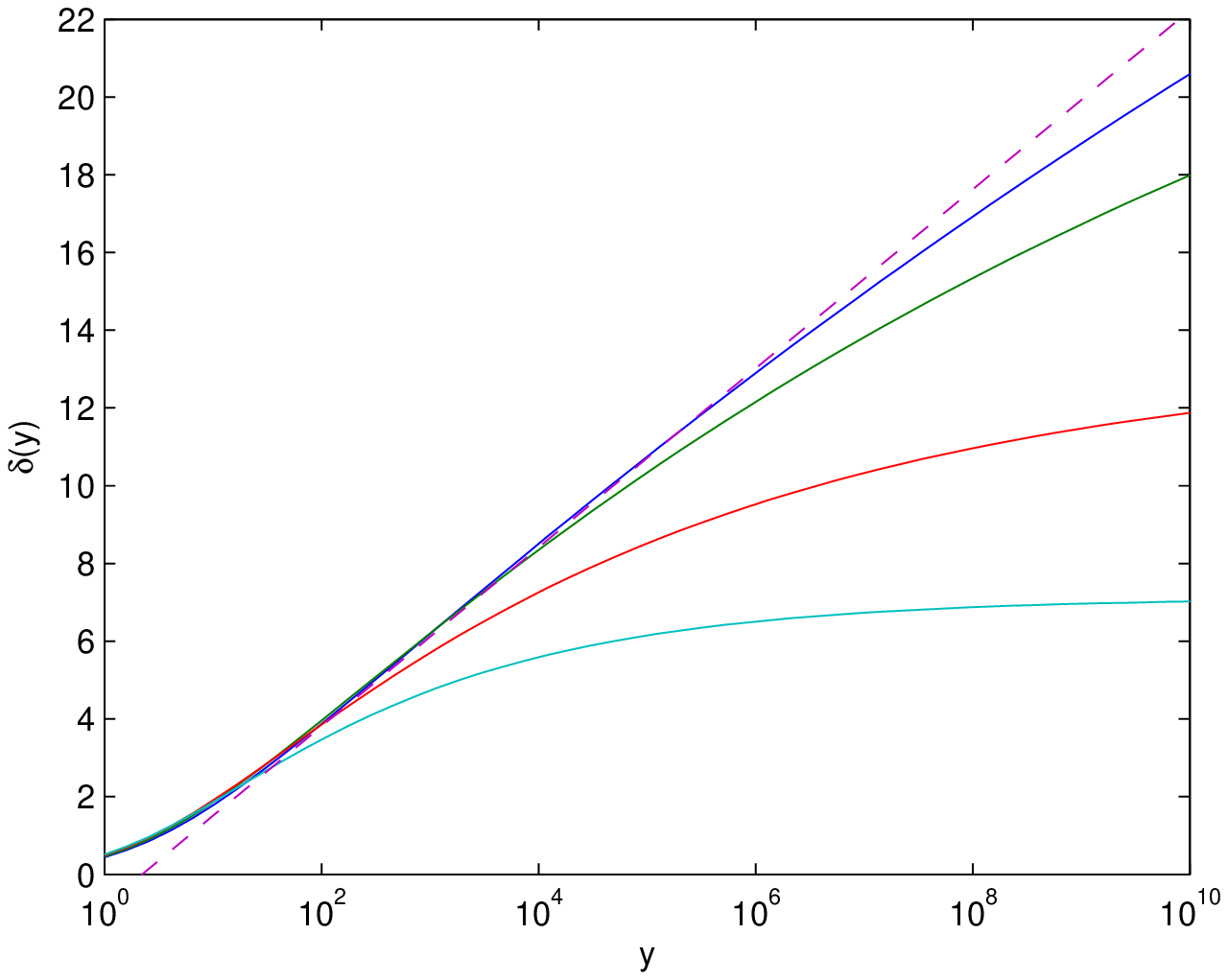}}} {\bf
Fig.~8:} \small{Dependence of $\delta(y)$ given by (\ref{d}) 
as a function of $y$ in logarithmic scale
for $\varphi=0.01, 0.02, 0.05, 0.1$ (top to bottom).
The straight dashed line corresponds to the logarithmic dependence $\ln y - 0.8$.
Not surprisingly, the closer $\varphi$ is to $0$, the larger is the range 
of $\ln y$ over which the approximation $\delta(y) \sim \ln y$ holds.
}
\end{quote}

\clearpage

\begin{quote}
\centerline{\includegraphics[width=14cm]{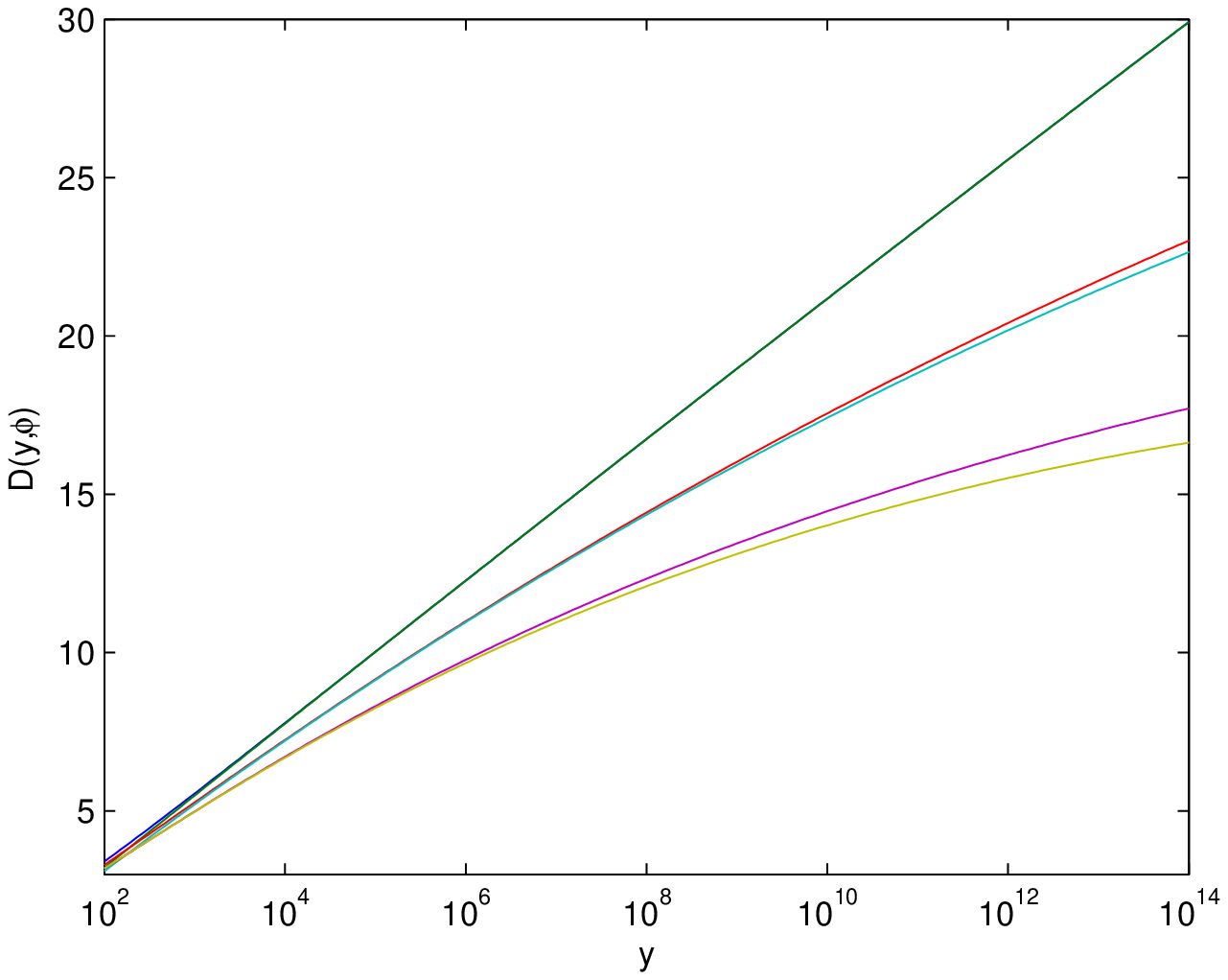}}
{\bf Fig.~9:} \small{Plots of the exact function $D(y,\varphi)$ and its
approximation (\ref{eqq7}) for $\varphi=0.001, 0.01, 0.02$ (top to bottom)}
\end{quote}


\begin{thebibliography}{}
\bibitem{M2} Mandelbrot, B.B., Multiplications al\'eatoires it\'er\'ees et distributions
invariantes par moyenne pond\'er\'ee al\'eatoire, I \& II. Comptes Rendus
(Paris): 278A, 289-292 \& 355-358.
\bibitem{citation} For the sake of
conciseness, we do not show the corresponding figures which are 
available upon request from the authors.
\end{thebibliography}
\end{document}